# Time Dependent Third Order Nonlinear Optical Susceptibility Tensor Components of the Carbon Disulfide Molecule using Femtosecond Degenerate Four-Wave Mixing

Prajal Chettri, and Shailesh Srivastava [*]

*Department of Physics, Sri Sathya Sai Institute of Higher Learning (SSSIHL), Andhra Pradesh, India-515134,*

*Corresponding author

E-mail address: shaileshsrivastava@sssihl.edu.in

## Abstract

*We report time resolved studies and measurement of all the components of the third-order nonlinear optical susceptibility of $CS_2$ at 515 nm using degenerate four wave mixing. The use of a pulse of width 450 fs, which is comparable to the time scale of the molecular reorientation dynamics of $CS_2$ makes our study unique. Given the wide use of this molecule as a reference material, our results will not only be a useful addition to the data base of the nonlinear optical properties of $CS_2$, but also provide additional insights into the time dynamics of the nonlinear processes involved. These results emphasize the importance of the time scales and wavelengths used to report the nonlinear optical susceptibility values for the reference material too.*

## 1. Introduction:

A wide variety of novel materials ranging from semiconductors to special glasses, conjugated organic molecules to nanoparticles, and even functionalized deoxyribonucleic acid (DNA) have been studied over the years for their third-order optical nonlinearities using high energy pulsed lasers for their potential applications in optical communication, optical computing, optical switching etc.[1]–[10]. There are number of approaches which can be used to characterize the third-order optical nonlinearities of these novel materials. These methods include the Z-scan technique [11], [12] and the degenerate four-wave-mixing (DFWM) method[13]–[16]. Since the nonlinear susceptibility is a complex quantity, it has both real part and imaginary parts. The real part gives information about the nonlinear refractive index and the imaginary part gives the information about the nonlinear absorption co-efficient of the material. DFWM relates to the absolute value of the nonlinear susceptibility i.e. $\left|\chi^{(3)}_{ijkl}\right|^2$.

DFWM is a process which results from the third order nonlinear optical interaction of light waves in a medium. It can provide information about the magnitude and temporal response of all the tensor components of the third-order nonlinear optical susceptibility. This process consists of three interacting beams of the same wavelength (degenerate) that include two pump beams and a probe beam. The strength of the fourth beam depends on a coupling constant that is proportional to the effective third order nonlinear susceptibility. When the interacting waves do not all have the same frequency, the process is said to be non-degenerate[17]. An example of this is Coherent Anti-Stokes Raman Spectroscopy (CARS)[18]. DFWM is most often achieved using the following geometries: a) Forward Box CARS or forward folded Box CARS geometry[2], [5], [19], b) Optical Phase Conjugate Geometry (backward geometry) [20], [21], and c) Two beam DWFM (forward geometry)[22], [23].

Since Carbon disulfide, has a relatively high third order nonlinear susceptibility, and is easily available, it is widely used as a standard reference for simplifying the measurement of third-order nonlinear susceptibilities[24]–[27]. It therefore becomes very essential to accurately characterize its nonlinear susceptibility. It is no surprise that over the last four decades, $CS_2$ has been the subject of many studies for its third order nonlinear optical properties, these findings have added to the database of nonlinear materials [6].

There can be various mechanisms responsible for the optical nonlinearities in a material, and some of the common mechanisms responsible are the instantaneous electronic response, molecular reorientation effects, as well as the electrostriction and thermal effects. Since electrostriction and thermal nonlinearities are relatively slow, these can be observed using nanosecond or longer duration laser pulses, or also using high repetition rates with femtosecond lasers. At the femtosecond time scales in $CS_2$, the dominant mechanisms involved are the instantaneous electronic response, and the slightly slower response due to molecular reorientations. The overall nonlinear response time of these molecules has been found to vary anywhere between 100 fs to 2 ps [28]. The reported values of the nonlinear susceptibilities have therefore been determined by the relative contributions of both these mechanisms. At the shortest time scales, the dominant contribution to the nonlinearity of the molecules is the instantaneous electronic response, while molecular reorientation is the dominant mechanism for slightly longer pulse widths. Many time domain studies have also been conducted on $CS_2$ in the past. These studies however focused on the electronic response of the molecules because of the timescales involved [27], [29], [30]. M. Reichert et al. [31] used a beam deflection technique to measure the third-order response of $CS_2$ over a broad temporal range using a 50 fs Ti:sapphire laser source at 800 nm. They further validated their technique using DFWM measurements using a 42 fs, Ti:sapphire laser source at 700 nm. The third-order nonlinear susceptibility value of $CS_2$ obtained by them was $(1.3 \pm 0.3) \times 10^{-13}$ esu. X. Q. Yan et al. [32] reported it to be $(9.8 \pm 0.7) \times 10^{-14}$ esu using a 125 fs Ti:sapphire laser source at 800 nm.

In this study, we used a pulse width of 450 fs, at 1030 nm. Since the pulse width is comparable to the molecular reorientation response time of $CS_2$, the relative contributions due to both electronic and molecular reorientation change during the interaction. Polarization dependence of the molecular reorientation effects further changes things. This makes our results different from the studies done earlier. We performed time dependent studies using various polarizations over a broad temporal delay range. Our studies provide additional insights into the molecular reorientation dynamics and add to the database of the nonlinear properties of $CS_2$. Our results would help researchers to compare the magnitudes of nonlinearities of novel materials, especially when working with similar pulse widths and wavelengths. Interestingly, the magnitude of the third-order nonlinear optical susceptibility of $CS_2$ obtained in our study is about ten times higher than the previously reported values in [31], [32].

## 2. Theory:

*2.1 Time-resolved Transient Grating technique*

To study the re-orientational dynamics of $CS_2$ molecules, the ultrafast time-resolved Transient Grating (TG) technique is widely used. This technique has a high signal-to-noise ratio and is often used for the Four-wave mixing process. Here, two pump pulses cross each other in the sample, leading to the formation of interference fringes in the region of overlap. If the sample has any light intensity dependent property, like intensity dependent refractive index, or nonlinear absorption, then its response imitates the interference pattern, leading to the periodic variation of refractive index. Thus an optical grating is created in the sample. For liquids, this grating formation is due to the third order nonlinear susceptibility, and could be created due to any of the mechanisms mentioned in the introduction. The amplitude of the periodic refractive index variations is directly proportional to the third order nonlinear susceptibility.

A low intensity probe pulse, with a variable delay, crosses the sample and gets partially diffracted by these gratings. The diffraction efficiency is proportional to the square of the grating amplitude. The nonlinearity induced grating relaxes back at a time scale that is dependent on the nonlinear process responsible for it. Since the probe beam is delayed with respect to the pumps, it responds to the amplitude of the grating at the delayed instant. The intensity of the diffracted beam therefore gives information of the temporal dynamics of the nonlinear response of the medium.

Phase matching of the wave vectors of the interacting beams should be satisfied in a DFWM experiment as it is directly related to the strength of the diffracted beam. The phase matching condition can be represented as:

$$\boldsymbol{k}_{dif} = \boldsymbol{k}_{pr} + \boldsymbol{k}_{pu1} - \boldsymbol{k}_{pu2} \quad (1)$$

Where $\boldsymbol{k}$ stands for the wave vector and the subscripts dif, pr, *pu1, pu2* denote the diffracted beam, the probe beam and the two pump beams respectively. The diffracted intensity is given by the following relation [33]:

$$I_{dif}(\omega_{dif}) = \left(\frac{2\pi}{nc}\right)^4 \cdot \frac{d^2 \omega_{dif}^2}{\varepsilon(\omega_{dif})} \cdot \left|\chi_{ijkl}^{(3)}\right|^2 I_{pr}(\omega_{pr}) \cdot I_{pu1}(\omega_{pu1}) \cdot I_{pu2}(\omega_{pu2}) \cdot sinc^2\left(\Delta k \cdot \frac{d}{2}\right) \quad (2)$$

Here, $\left|\chi_{ijkl}^{(3)}\right|$ is the magnitude of the third-order nonlinear susceptibility. $I_{pr}$, $I_{pu1}$ and $I_{pu2}$ are the intensities of the input probe and the two pump beams, n is the refractive index of the material, $\omega_{dif}$ is the frequency of the diffracted beam, $\varepsilon(\omega_{dif})$ is the dielectric constant of the nonlinear material, c is the speed of light in vacuum, and *d* the interaction length. Δk is the magnitude of the phase mismatch and is given by:

$$\Delta k = k_{pr} + k_{pu1} - k_{pu2} - k_{dif} \quad (3)$$

Due to the $sinc^2\left(\Delta k \cdot \frac{d}{2}\right)$ dependence the intensity of the diffracted beam decays periodically with the interaction length, if the phase matching condition is not satisfied (Δk ≠ 0). The forward folded box CARS geometry of DFWM makes it easy to achieve (Δk = 0) and make this term unity.

In the case of a time dependent observation, the diffracted intensity at the probing time $t_d$ is given as: [33]

$$I_{dif}(t_d) = \left(\frac{2\pi}{nc}\right)^4 \cdot \frac{d^2 \omega_{dif}^2}{\varepsilon(\omega_{dif})} \cdot \int_{-\infty}^{+\infty} dt \cdot I_{pr}(t_d - t) \cdot \left[\int_{-\infty}^{t} d\tau \cdot F_{ijkl}^{(3)}(t - \tau) \cdot I_{pu}(\tau)\right]^2 \quad (4)$$

Where, the two integrals represent the convolution of the sample's nonlinear response with both pump pulses (last term) and with the probe pulse (first integral). Here $F_{ijkl}^{(3)}$ is the Fourier transform of $\chi_{ijkl}^{(3)}$ and represents the third order nonlinear optical response function of the medium. $F_{ijkl}^{(3)}$ is therefore also referred to as the time dependent nonlinear optical susceptibility [38]. Eq.4 actually assumes equal pump intensities, and in case of unequal intensities the second integral squared will be replaced by a product of two such integrals with $I_{pu1}(\tau)$ and $I_{pu2}(\tau)$.

When all the interacting fields are much shorter than the nonlinear response times involved, they can be considered as delta functions. In this case the convolution integrals collapse to $\left|F^{(3)}_{ijkl}(t_d)\right|^2 I_{pr}I_{pu1}I_{pu2}$ and we obtain the instantaneous time domain nonlinear optical susceptibility at the delayed time. On the other hand when all the fields are monochromatic continuous waves then one uses the frequency domain equation (2), since the power measurements are directly at $\omega_{dif}$.

When using transform limited nano or picoseconds pulses the fields can still be considered monochromatic and the time domain power measurements at zero delay can directly use equation (2). The ratio of the pulse energy and pulse width gives the peak powers, while the focused spot sizes are used to calculate the peak intensities. This is what many earlier studies of the third order susceptibility measurements for $CS_2$ report [6]. In case one uses transform limited Gaussian pulse widths of about 500 fs, the spectral bandwidth at 515 nm would be about 0.9 nm, and the time domain power measurements at zero delay would then relate to $\chi^{(3)}_{ijkl}$ averaged over the 0.9 nm spectral bandwidth in equation (2). Studies using less than 50 fs pulses with spectral bandwidths more than 10 nm, still tend to use equation (2) to calculate the third order nonlinear optical susceptibility at the center wavelength, often without mentioning the effective spectral bandwidth over which this is averaged. This could make a difference if there is large dispersion of $\chi^{(3)}_{ijkl}$ at the wavelength.

In our case the pulses are not transform-limited and the spectral bandwidth is about 8 nm, so when using equation (2) we remind that we are reporting $\chi^{(3)}_{ijkl}$ at 515 nm averaged over $\pm$ 4nm. The dispersion of $\chi^{(3)}_{ijkl}$ in this wavelength range is neglected in our results. Our results do not give $F^{(3)}_{ijkl}(t_d)$ since the interacting pulse widths (470 fs) cannot be approximated to delta functions.

We remind ourselves that $\chi^{(3)}_{ijkl}(\omega_4 = \omega_3+\omega_2+\omega_1)$ is the nonlinear susceptibility, where the subscripts i, j, k and l indicate the electric field polarization directions, and correspond to the frequencies, $\omega_4$, $\omega_3$, $\omega_2$ and $\omega_1$ respectively. For isotropic media like liquids, only 4 of the 81 components of the $\chi^{(3)}_{ijkl}$ are nonzero and are related by $\chi^{(3)}_{1111} = \chi^{(3)}_{1122}+\chi^{(3)}_{1212}+\chi^{(3)}_{1221}$. Here conventionally the subscripts 1 & 2 represent any two perpendicular polarizations, and could mean x, y or z directions due to the isotropic nature of the medium. For the intensity dependent refractive index in degenerate four-wave mixing the relevant nonlinear susceptibility is $\chi^{(3)}_{ijkl}(\omega = \omega+\omega-\omega)$. Since the first and second interacting frequencies are the same, the property of intrinsic permutation symmetry makes $\chi^{(3)}_{1122}=\chi^{(3)}_{1212}$. This leads to the relation $\chi^{(3)}_{1111} = 2\chi^{(3)}_{1122}+\chi^{(3)}_{1221}$. Each of these $\chi^{(3)}_{ijkl}$ represents the sum of the contributions due to all the mechanisms responsible for the nonlinearities.

*2.2 Forward folded-BOX CARS geometry*

The folded-BOX CARS geometry[2], [19] is shown in Figure 1. In this geometry, three input beams are arranged in such a way that, they pass through three corners of a square. Under these conditions, the diffracted fourth beam, which is generated in the phase matched direction, passes exactly through the fourth corner as shown in the Figure 1. The advantages of using this particular geometry are well known. In particular, it is easier to achieve phase-matching. Also since the diffracted signal beam is spatially separated it minimizes background noise from the other beams and makes it convenient to place the detector.

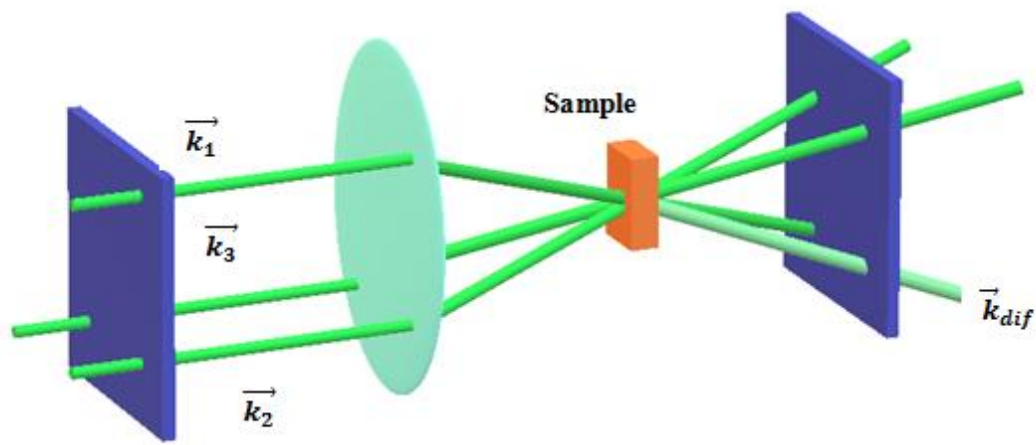

Figure 1: 3D view of folded Box-CARS geometry

## 3. Methods:

*3.1 Experimental Set-up*

Figure 2 displays the experimental configuration. The 515 nm, input beam was the second harmonic of the output from a Yb doped fiber laser system (Menlo Systems BlueCut-270 fs, 1030 nm, at 1 MHz rep rate). The SHG at 515 nm had a pulse width broadened to 470 fs due to group velocity mismatch, while propagating through a 2 mm BBO crystal, Type-1 phase matched.

The main SHG beam was divided into two pump beams using a 50-50 beam-splitter. One beam was directly used as one of the pumps (pump 2) and the other beam was further divided by an 80-20 beam-splitter. Due to the nature and limitations of the optical components that we had, the two pumps were not of equal intensities (Table 2). In the nonlinear susceptibility formulism (equation (2)) of the TG technique this will not affect the results. Though experimentally for a given total input power, if we divide into two equal pumps, one would get the largest product of $I_{pu1}I_{pu2}$, and the diffracted signal would be easier to detect.

The two pumps and the probe were made parallel to each other, with a spacing of about 12 mm between them, at the three corners of a square. A convex lens of 50 cm focal length was used to converge the three beams to its focal point. The alignment for the zero path difference position was done by observing the three focused spots on a webcam, and iteratively adjusting the path lengths of the optical delays to maximize the contrast of the interference pattern. The sample was put in a quartz cuvette of length 1 cm and placed such that the focal spots were at the center of the cuvette. Fine alignment was done after placing the sample and maximizing the power of the diffracted beam.

All the power measurements were made using a power meter (Ophir Nova II) with a silicon photodiode sensor of dimensions 10 x 10 mm. An appropriate beam block with an aperture was used before the sensor to allow only the diffracted signal. The powers of the input beams used in this study are mentioned in Table 2.

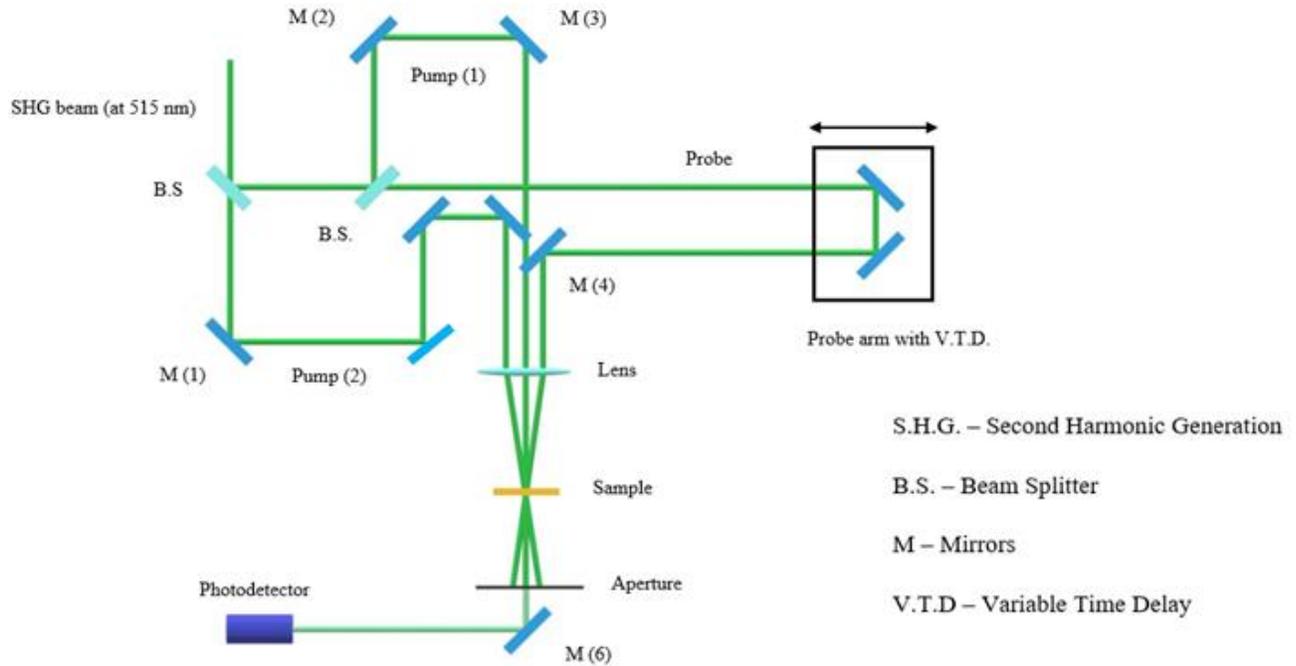

Figure 2: Experimental configuration for DFWM

*3.2 Time resolved studies*

For the time-resolved experiments, the probe beam and the pump 2 were kept on delay lines created using a micrometer translational stage and mirrors. All beams had vertical polarizations initially. A zero order half wave plate at 515 nm (Thorlabs), kept in any arm with its axis at 45° to the vertical produced horizontal polarization. For $\chi^{(3)}_{1111}$ measurements, all the beams had the same vertical polarization. The half wave plate was kept with its axis along the vertical for these measurements. For $\chi^{(3)}_{1122}$ measurements, the half-wave plate was kept in the path of the probe beam and the polarization was made horizontal. Finally for $\chi^{(3)}_{1221}$ measurements, the half-wave plate was kept in the path of the pump 2 to produce the horizontal polarization. Even though the half wave was not required for the $\chi^{(3)}_{1111}$ measurements, it was appropriately placed in one arm for all measurements, so that the net intensity product in equation (2) would remain unchanged. The path lengths were fine tuned in each case to correctly obtain the zero delay position in the presence of the wave plate.

# 4. Results and Discussions:

*4.1 Polarization studies:*

As mentioned before, the polarization of any of the three input beams could be changed by placing a half wave plate whose axis was 45° with respect to the vertical direction. From the tensor nature of third order susceptibility of isotropic media, one can predict the polarization of the DFWM signal [34]. The diffracted beam's polarization was in agreement with the theoretically expected output polarization in each case (Table 1).

| Pump 1 | Pump 2 | Probe | DFWM signal (expected) | DFWM signal (observed) |
|---|---|---|---|---|
| x | x | x | x | x |
| x | x | y | y | y |
| x | y | y | x | x |

Table 1: Polarization studies of the input versus the output beams

*4.2 Third-order nonlinear susceptibility measurements of $CS_2$ molecules:*

Third-order nonlinear susceptibilities were measured for various input beam polarizations and these are tabulated in Table2. The power measurements were made by averaging over 30 seconds for each reading. The error values indicate the variation after 30 seconds of averaging. The fluctuations in the diffracted signal were attributed mainly to the small pump fluctuations, some unavoidable background scatter and some random occasional turbulence induced in the $CS_2$ . The percentage fluctuations on the power meter were made small by the large averaging time. The background scatter that existed due to the pumps in the absence of the probe was also averaged and subtracted to obtain each reading. This made all readings highly precise. There was however no way to check the accuracy of the final results, as no reports existed in literature using these time scales and wavelength.

| Polarizations | Pump 1 (mW) | Pump 2 (mW) | Probe (mW) | DFWM signal ($I_{dif}$) (µW) | Effective $\chi^{(3)}$ at zero delay ($10^{-12}$) (in e.s.u) | % error in $\chi^{(3)}$ |
|---|---|---|---|---|---|---|
| 1111 | 16.7±0.1 | 43.2±0.1 | 1.56±0.01 | 130.3±5 | (5.81±0.11) | 2 |
| 1122 | 21.6±0.1 | 55±0.1 | 1.79±0.01 | 21.3±0.8 | (1.71±0.03) | 1.75 |
| 1221 | 16.9±0.1 | 44.3±0.1 | 1.58±0.01 | 40.1±1.5 | (3.14±0.06) | 1.9 |

Table 2: Third-order nonlinear susceptibility measurements of $CS_2$ molecules for various input beam polarizations.

In Table 2, 1111 corresponds to all the incident beams being parallel in polarization, 1122 corresponds to the probe being perpendicular in polarization with respect to the two pumps and 1221 corresponds to one of the pumps being perpendicular in polarization with respect to the other pump and probe.

The third-order nonlinear susceptibility $\chi^{(3)}_{1111}$ for $CS_2$ was calculated to be $(5.81\pm0.11)\times10^{-12}$ esu. We reiterate that to the best of our knowledge, the third order nonlinear optical susceptibility values at this time scales and wavelength have not been reported for $CS_2$. Most studies in literature are at the nanosecond or few picoseconds timescales. The reports on the femtosecond time scales seem to be at the 100 fs or shorter time scales and at 800 nm wavelength, using the Ti-Sapphire laser

*4.5 Time dependent studies:*

The strength of the diffracted signal was observed with increase in time delay of the probe. This was repeated for various polarization states of the probe and the pumps, corresponding to $\chi^{(3)}_{1111}$, $\chi^{(3)}_{1122}$ and $\chi^{(3)}_{1221}$. These results are plotted in figure 4. On performing multiple trials, as well as the 30 seconds averaging, all observations were found to be repeatable to within about 4% error, so no separate error bars are put in the plots. In Table 2 the column on DFWM signal also shows these errors as power values.

Theoretically [34], the change in refractive index due to molecular reorientation is proportional to $\langle\cos^2(\theta)\rangle-1/3$, where $\theta$ is the angle between the probes electric field direction and the molecular axis of symmetry. The angular brackets $\langle\ \rangle$ denote an ensemble average over all the possible solid angles. In our experiment, due to the pumps, the molecular axis is forced to orient along the pump's electric field direction, so $\theta$ is also the angle between the probe and the pumps.

When $\theta=54.73°$ the $\langle\cos^2(\theta)\rangle-1/3 = 0$, and the contribution of nonlinearity due to molecular reorientation is eliminated. This angle is therefore popularly known as the 'magic angle'.

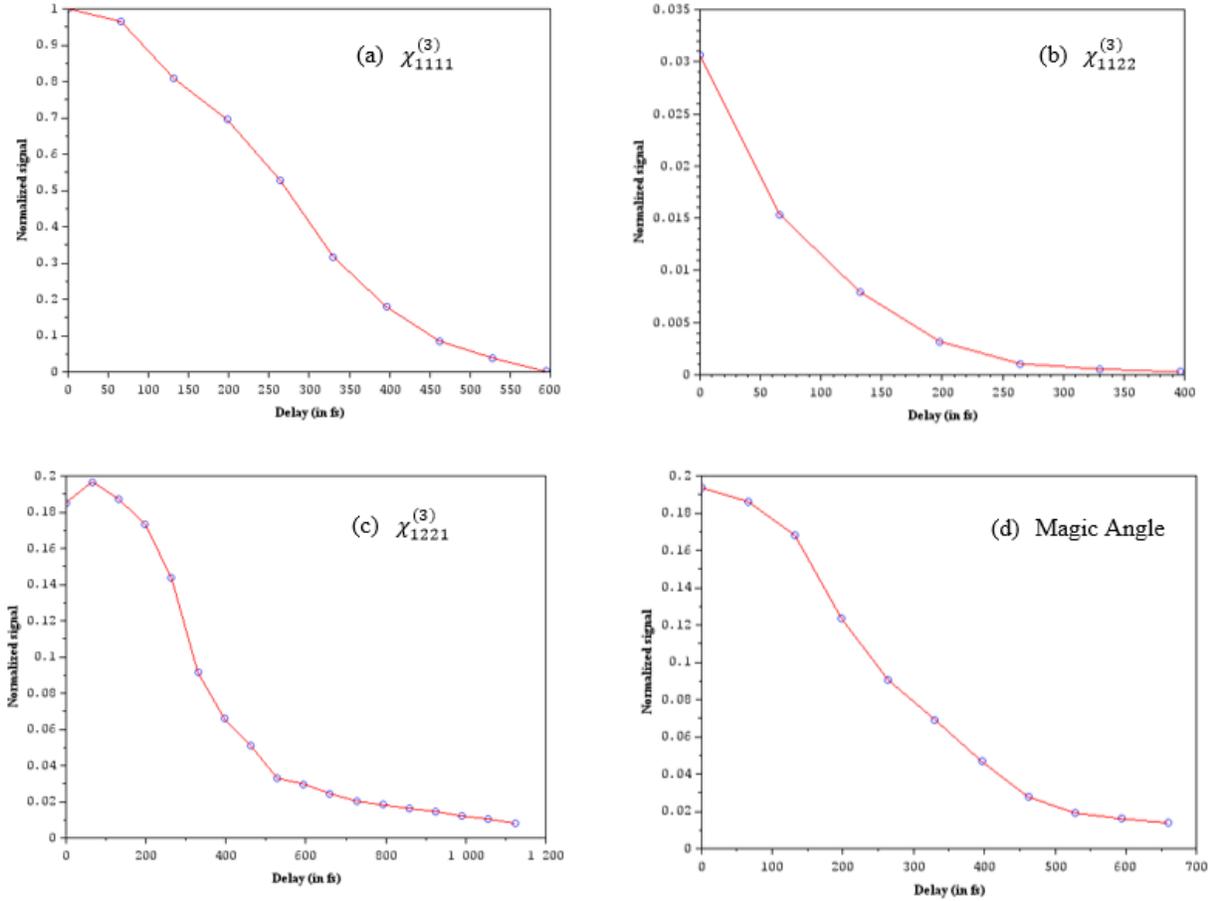

Figure: 4(a) Plot for $\chi^{(3)}_{1111}$ when all the input beams were parallel in polarization; 4(b) Plot for $\chi^{(3)}_{1122}$ when the probe was perpendicular in polarization with respect to the pumps; 4(c) Plot for $\chi^{(3)}_{1221}$ when one of the pumps was perpendicular in polarization with respect to the other pump and the probe; 4(d) Plot at magic angle i.e. at $54.73^0$

In the first graph (Figure 4(a)) where all the incident beams were parallel in polarization, we expect both electronic and molecular contributions to be at maximum due to $\chi^{(3)}_{1111}$ as well as due to $\theta = 0$. The diffracted intensity in this case at zero time delay was found to be at the maximum as expected. The peak of this graph, $I_{dif1111}$, was kept at the maximum normalized value = 1. All other observations were normalized with respect to this value.

In the second graph (Figure 4(b)), the probe is perpendicular with respect to the two pumps. The relevant susceptibility is now $\chi_{1122}$. In this case the probe beam sees molecules oriented perpendicular to its electric field polarization. As the orientation angle of the molecules, $\theta$, with respect to the probe is $90^0$, we find that the change in refractive index of the probe due to molecular reorientation is initially negative since the $\langle \cos^2(\theta) \rangle$ term averages to zero. So, the contribution due to molecular reorientation (negative) opposes the non-resonant electronic change (positive). The electronic contribution to the signal should be about 1/9 of the all-parallel value since $\chi_{1122} = (1/3)\,\chi_{1111}$ and the diffracted signal is proportional to square of $\chi_{1122}$. We therefore expect a signal much smaller than about 1/9 of the all-parallel value, with both these opposing effects acting together. This explains the maximum of 0.031 at zero delay in this case.

As the pump is no longer present after about 450 fs, the CS$_2$ molecules freely return back to random orientations within their own characteristic relaxation time of 1-2 ps. Interestingly the ensemble average of the $\cos^2(\theta)$ term over all solid angles in the absence of strong field [34] is also $\langle \cos^2(\theta) \rangle =$ 1/3. This makes the final contribution due to molecular reorientation relax towards zero from the initial negative value. Since the change in refractive index values due to molecular reorientation remains negative throughout the interaction, the effective decay rate of the combined nonlinear electronic and reorientation response is much faster than the molecular relaxation time. That explains the difference in the shape of figure (4b)

In the third graph corresponding to $\chi^{(3)}_{1221}$ (Figure 4(c)), one of the pumps is perpendicular with respect to the probe and the other pump. Due to the mutually perpendicular pumps no interference fringes are formed, so no intensity variations occur along the length. Since the sample refractive index is slightly different for the two pump orientations, a phase difference develops between the pump fields, and this periodically varies the state of polarization along the length. This periodic polarisation state variation makes the molecular orientations also vary periodically. The final result is an orientational grating [33]. If the time required for this orientational grating to fully form is a bit longer and within our experimental time resolution, the diffracted signal should initially increase and maximize, and then go to zero when the molecules relax to the isotropic orientation.

Figure 4(d) shows the result for the probe oriented at the magic angle. In this case the probe can be split into two components. The component parallel to the pump would generate a $\chi^{(3)}_{1111e}$ type response and the component perpendicular to the pump would generate a $\chi^{(3)}_{1122e}$ type of response. The 'e' subscript is to indicate that only electronic components contribute at the magic angle. The intensity of the probe parallel to the pump at the *Magic angle* is roughly reduced to $\sim 0.33 = \cos^2(54.73)$, i.e. 33% of the total probe intensity and so the perpendicular component is reduced to 67% of the total probe intensity. Since the diffracted intensity is linearly dependent on the probe intensity, it should proportionately reduce the diffracted output at magic angle (MA) as indicated below

$$I_{difMA} = const \times \left\{ 0.33 \left|\chi^{(3)}_{1111e}\right|^2 + 0.67 \left|\chi^{(3)}_{1122e}\right|^2 \right\} =$$

$$= const \times \left\{ 0.33 \left|\chi^{(3)}_{1111e}\right|^2 + 0.67 * (1/3)2 \left|\chi^{(3)}_{1111e}\right|^2 \right\} = 0.404 const \times \left|\chi^{(3)}_{1111e}\right|^2.$$

In figure 4a however the output is

$$I_{dif1111} \propto const \left|\chi^{(3)}_{1111}\right|^2 = \propto const \left\{ \left|\chi^{(3)}_{1111e}\right|^2 + \left|\chi^{(3)}_{1111M}\right|^2 + 2\chi^{(3)}_{1111e}\chi^{(3)}_{1111M} \right\} \text{ (assuming all real } \chi \text{)}$$

If the contribution in Figure 4 (a) is due to both electronic and molecular reorientation, we must obtain in figure 4 (d), a normalized value much lesser than $0.404 I_{dif1111}$. Our data for the case of the *Magic angle* shows a maximum of 0.19 $I_{dif1111}$. This proves the presence of molecular reorientation at zero delay for the all-parallel case.

The nonlinear optical response at the *Magic angle* should be a delta-function (due to the ultrafast electronic response acting all alone). The theoretical fit produced using a Gaussian pump pulse of FWHM 450 fs, and a delta function impulse response $F^{(3)}_{ijkl}$ in Eq.4 was plotted along with the experimentally observed results for the *Magic angle* (54.73⁰). The fit reaffirms the fact that only instantaneous nonlinear electronic response contributed at the *Magic angle* as expected, and is a confirmation of the preceding discussion.

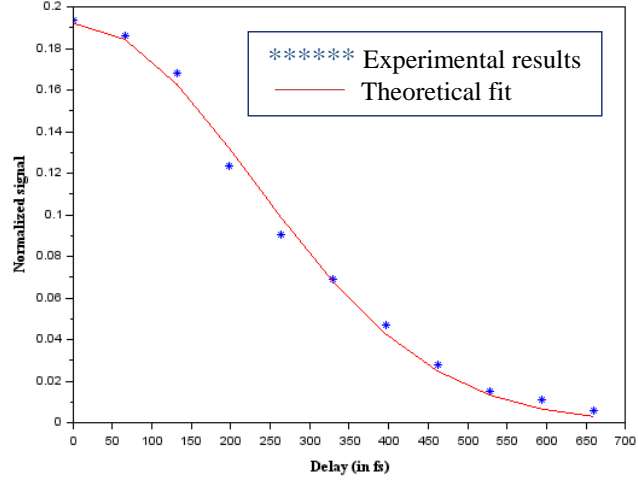

Figure: 5: Plot at magic angle ($54.73^0$) with its theoretical fit

### 4.3 Ruling out Thermal effects:

As mentioned in the introduction Thermal nonlinear effects can also contribute in a DFWM process. When bright and dark fringes are formed due to the two pumps interfering, it can create thermal gratings, if the medium absorbs due to either two-photon absorption, or even linear absorption. Since thermal effects require time to dissipate and generate, if these effects last as long as the inter-pulse separation time the effects can accumulate. Thermal effects thus become important at high repetition rates [39]. Also even if these effects do not accumulate they would at least last for pico to nano seconds. Since we did not see any diffracted signal after probe delays of 1-2 ps it implies that thermo-optic nonlinear effects were not present. In fact a look at the figure 4(b) shows that the signal vanishes above a time delay of about 450 fs. This is in spite of the fact that both pumps have the same polarization and interference fringes should be formed. Figure 4 (c) is for the case where both pumps are perpendicular ($\chi^{(3)}_{1221}$), so no interference fringes should be formed, and no thermal gratings. The fact that in this case, the diffracted signal lasts a little longer (800fs -1 ps) than the case with $\chi^{(3)}_{1122}$, conclusively rules out any contribution from thermal nonlinear effects. The absence of any absorption effects also tells that the $\chi^{(3)}_{ijkl}$ are all real in our experiment at 515 nm.

### 4.4 Non-Linear Ratios: [34], [36]

Theoretically, the ratios of the nonlinear susceptibilities for the electronic and molecular re-orientation response are given below:

Electronic response: $\dfrac{\chi^{(3)}_{1221}}{\chi^{(3)}_{1122}} = 1$ (5)

Molecular reorientation: $\dfrac{\chi^{(3)}_{1221}}{\chi^{(3)}_{1122}} = 6$ (6)

Experimentally, the overall effective nonlinear ratio was observed to be:

$$\frac{\chi^{(3)}_{1221}}{\chi^{(3)}_{1122}} = 1.84 \qquad (7)$$

This result confirms that $CS_2$ has both electronic and molecular response, since the ratio lies between 1 and 6. The ratio is similar to the one reported in [32] using 125 fs pulses at 800 nm. Time dynamic studies were however not done in that work. The value of 1.84 also indicates that we are probably closer to the non-resonant electronic ratio, rather than the ratio for molecular reorientation. This could possibly be explained because at the intensities and time scales involved all molecules do not align to the field direction [34], while all the molecules contribute to the non-resonant electronic response irrespective of the intensities of the beams.

From table 2, we see that $\chi^{(3)}_{1111} = 2\chi^{(3)}_{1122} + \chi^{(3)}_{1221}$ is only approximately satisfied within the limits of experimental errors. Some deviation from the equality is also because the relationship assumes an isotropic medium. However, in the presence of a linearly polarized pump, some molecules align in one direction and introduce anisotropy in the liquid.

Further, since we actually measure $\left|\chi^{(3)}_{electronic} + \chi^{(3)}_{Molecular}\right|^2$, if the different $\chi^{(3)}$ were complex and had imaginary parts, and if each of the processes had not yet dephased, then the two contributions would interfere and the net result would depend on the averaged phase difference during the interaction. Since all $\chi^{(3)}$ were real in our experiment, we measure signals proportional to the square of algebraic sum of the two contributions.

For the particular case of $\chi^{(3)}_{1122M}$ as explained earlier, the change in refractive index due to molecular reorientation is negative. So the diffracted signal that we measure is proportional to $\left|\chi^{(3)}_{1122e} - \chi^{(3)}_{1122M}\right|^2$

Therefore the ratio, $\frac{\chi^{(3)}_{1221}}{\chi^{(3)}_{1122}} = \frac{\chi^{(3)}_{1221e} \pm \chi^{(3)}_{1221M}}{\chi^{(3)}_{1122e} \pm \chi^{(3)}_{1122M}}$ at zero delay can give different results, due to the possibility of negative signs whose value can change during the time of integration over the pulse width. The same is true, if the susceptibilities are complex due to the phasor additions of the individual contributions. Under these conditions it makes interpretation of the ratio more difficult. One should to be aware of these details while reporting the susceptibility values or ratios whenever two or more nonlinear processes are involved.

## 5. Conclusion

In conclusion, we have investigated the degenerate four-wave mixing (DFWM) process with the folded-Box CARS geometry in carbon disulfide ($CS_2$) using a 450 femtosecond pulsed laser at 515 nm. All tensor components of the third-order nonlinear susceptibility were calculated, and $\chi^{(3)}_{1111}$ measured to be $(5.81 \pm 0.11) \times 10^{-12}$ esu. The time dependent studies and the nonlinear ratios at our timescale proved that the contribution to the nonlinearity at this time scale is due to both the instantaneous electronic response and the molecular reorientation, the former being more dominant. We believe to the best of our knowledge that the time dynamics of the nonlinear susceptibilities of $CS_2$ at these time scales, and wavelength have not yet been reported. Additionally, such insights into the time dependent relative contributions of two different mechanisms have been provided for the first time.